\documentclass[11pt, oneside]{article}   	

\usepackage{jheppub}

\usepackage{amsmath}
\usepackage{amssymb}
\usepackage{amsfonts}
\usepackage{amsthm}
\usepackage{amsbsy}
\usepackage{array}

\usepackage{mathtools}

\usepackage{tikz}

\usepackage{subcaption}

\usepackage{dsdshorthand}

\usepackage{graphicx}

\usepackage[vcentermath]{youngtab}

\newcommand\vol{\mathop{\mathrm{vol}}}
\newcommand\Disc{\mathrm{Disc}}

\newcommand{\la}{\langle}
\newcommand{\ra}{\rangle}

\renewcommand\bx{\mathbf{x}}

\newcommand*\link[1]{\hspace*{0em plus 1fill}\makebox{#1}}

\definecolor{darkgreen}{rgb}{0, 0.6, 0}

\makeatletter
\def\@fpheader{\ }
\makeatother

\title{
Holographic Correlators at Finite Temperature
}

\author{Luis F. Alday$^1$, Murat Kolo\u{g}lu$^1$, and Alexander Zhiboedov$^2$}
\affiliation{${}^1$Mathematical Institute, University of Oxford, Oxford, OX2 6GG, UK\\
${}^2$CERN, Theoretical Physics Department, 1211 Geneva 23, Switzerland
}

\date{}
\abstract{We consider weakly-coupled QFT in AdS at finite temperature. We compute the holographic thermal two-point function of scalar operators in the boundary theory. We present analytic expressions for leading corrections due to local quartic interactions in the bulk, with an arbitrary number of derivatives and for any number of spacetime dimensions. The solutions are fixed by judiciously picking an ansatz and imposing consistency conditions. The conditions include analyticity properties, consistency with the operator product expansion, and the Kubo-Martin-Schwinger condition. For the case without any derivatives we show agreement with an explicit diagrammatic computation. The structure of the answer is suggestive of a thermal Mellin amplitude. Additionally, we derive a simple dispersion relation for thermal two-point functions which reconstructs the function from its discontinuity.}
\preprint{
\link{CERN-TH-2020-155}}

\begin{document}

\maketitle
\pagenumbering{roman}
\setcounter{page}{2}
\newpage
\pagenumbering{arabic}
\setcounter{page}{1}

\section{Introduction}
\label{sec:introduction}

Finite temperature dynamics of interacting systems is of great interest both theoretically and experimentally. Thermal observables contain a plethora of dynamical information and are typically very hard to compute.\footnote{In higher dimensions, the standard examples include perturbative theories, see e.g.~\cite{Kapusta:2006pm,Katz:2014rla},  and holographic computations, see e.g.~\cite{Hartnoll:2016apf}. One can also compute certain thermal two-point functions analytically in the large-$q$ limit of the SYK chain~\cite{Gabor2020}.} Consider for example a thermal two-point correlation function. At large frequencies and momenta it is controlled by the standard Euclidean OPE, see e.g.~\cite{CaronHuot:2009ns}. At low frequencies and momenta, however, it describes hydrodynamical transport properties of the system \cite{martin1963leo}. The interpolation between the two regimes necessarily comes with a scale, the temperature $T$, which breaks symmetries that render the dynamics exactly solvable, such as integrability or supersymmetry.\footnote{Applying conformal field theory bootstrap techniques at finite temperature $T$ is challenging as well. One significant reason is the absence of the non-negativity of thermal one-point functions, see~\cite{ElShowk:2011ag,Iliesiu:2018fao} for a detailed discussion.} Because of this having some exactly solvable examples of thermal observables in QFTs can potentially be very useful. Such examples can serve as a playground to test and develop theoretical ideas that can then be hopefully adapted to more complicated and realistic situations. It is the primary purpose of this paper to construct a new class of such examples.

In this paper, we study quantum field theory in $AdS_{d+1}$ at finite temperature $T$. We consider a massive real scalar field $\Phi$ with a nontrivial quartic interaction
\be
\label{eq:perturbative QFT in AdS}
S_{E} =\int d^{d+1} x \sqrt{g} \left( {1 \over 2} \Phi (- \nabla^2 + m^2) \Phi + {\lambda_0 \over 4!} \Phi^4 + \sum_{k=2}^{L} \lambda_k \nabla^{2k} \Phi^4 \right) ,
\ee
where $g$ is the thermal Euclidean AdS metric and we kept indices on the derivative contractions implicit.\footnote{Corrections to the thermal partition function in this setup were recently considered in \cite{Kraus:2020nga}.} The field $\Phi$ is holographically dual to a real scalar primary operator $\phi$ on the boundary, of scaling dimension $\Delta$ given by $m^2 = \Delta(\Delta-d)$, where we set the AdS length to one. This theory defines a set of conformal correlation functions \be
\la \phi(x_1) ... \phi(x_n) \ra_{\beta}
\ee on the boundary of the AdS space as a series
expansion in $\lambda_i$, where $\beta \equiv {1 \over T}$. Since these correlators emerge naturally from a theory in AdS, we call them holographic correlators. The theory (\ref{eq:perturbative QFT in AdS}) has $\mathbb{Z}_2$ symmetry that acts as $\Phi \to - \Phi$ which implies that $\la \phi \ra_{\beta}=0$. In this paper we study the first non-trivial correlator of this type, namely $\la \phi(x_1)\phi(x_2) \ra_{\beta}$.

Note that in this case the theory in the bulk does not have dynamical gravity which corresponds to the fact that theory on the boundary does not have the stress-energy tensor. Generalizing the analysis of the present paper to this more interesting and complicated case is a task that we will not address here. 

The leading order correction to the four-point function $\la \phi(x_1) \phi(x_2)\phi(x_2) \phi(x_4) \ra$ at zero temperatures was famously analyzed in~\cite{Heemskerk:2009pn}. There, it was found that at zero temperature a direct connection between perturbative QFTs in AdS (\ref{eq:perturbative QFT in AdS}) and solutions to the conformal crossing equations of the correlator $\la \phi(x_1) \phi(x_2)\phi(x_2) \phi(x_4) \ra$ exists. This connection was furthermore understood as a one-to-one correspondence between the possible perturbations of the free QFT in AdS and the perturbative solutions to the crossing equations. 

In this paper we generalise the analysis of~\cite{Heemskerk:2009pn} to finite temperature $T>0$. We explicitly construct thermal two-point functions $\la \phi(x_1) \phi(x_2) \ra_{\beta}$ holographically dual to the same theory~(\ref{eq:perturbative QFT in AdS}) considered in~\cite{Heemskerk:2009pn}, albeit now at nonzero temperature. As in the $T=0$ case, our solutions are to leading order in the couplings $\lambda_i$. At finite temperature, the boundary geometry is given by $S_{\beta}^1 \times \mathbb{R}^{d-1}$, where the temperature sets the circumference $\b$ of the thermal circle, $S^1_\beta$. The role of the crossing equations for the thermal case is played by the Kubo-Martin-Schwinger (KMS) condition, which states that thermal correlators are periodic along the thermal circle. The KMS condition imposes nontrivial constraints on thermal correlators, which has been used to develop a thermal bootstrap problem for thermal two-point functions~\cite{ElShowk:2011ag,Iliesiu:2018fao}. Provided with the data of the $T=0$ solutions of~\cite{Heemskerk:2009pn}, we impose KMS and other consistency conditions on a carefully chosen ansatz, and are able to construct an infinite family of solutions to the thermal two-point function bootstrap problem.

Let us be more specific about our solutions, and provide an example. Free fields in thermal AdS are holographically dual to mean field theory (MFT) on the boundary $S_{\beta}^1 \times \mathbb{R}^{d-1}$. The thermal two-point function of the MFT field $\phi$ is one of the few exact known solutions to the thermal bootstrap. Picking coordinates $x=(\tau,{\bf x})\in S^{1}_\beta \times \R^{d-1}$, it takes the form
\be
\label{eq:mftfirst}
G_\De(x)\equiv\<\f(x)\f(0)\>_\b^{\text{MFT}} = \sum_{m=-\infty}^\infty \frac{1}{((\tau+m)^2 + {\bf x}^2)^\Delta}\,.
\ee 
Note that $G_\De(x)$ only converges for $\De>1/2$; accordingly, we will restrict to this case in our work. 
Turning on perturbative interactions~(\ref{eq:perturbative QFT in AdS}) in the bulk corrects the thermal correlators of the boundary MFT, and it is specifically the first order $\l_i^1$ corrections to $G_\De(x)$ that we seek to construct. By imposing consistency conditions on our ansatz, we find, for example, that turning on a quartic interaction $\l_0 \Phi^4$ in the bulk results in a leading order correction of the form
\be
\label{eq:correction from bulk quartic interaction}
\<\f(x)\f(0)\>_\b |_{\lambda_0^1} = \oint_{s_0-i\infty}^{s_0+i\infty} \frac{ds}{2\pi i} \Gamma(s)^2 \Gamma(\De-s)^2 M_{\beta}(s) G_s(x)\, , 
\ee where $ 0<{\rm Re} \ s_0 < \Delta - (d-1)/2$. The function $M_{\beta}(s)$ is entirely fixed by the corresponding $T=0$ solution of~\cite{Heemskerk:2009pn} --- namely by the order $\l_0$ anomalous dimensions $\gamma_0(n)\equiv \gamma_{n,\ell=0}$ of spin $\ell=0$ operators given explicitly in (\ref{eq:spin 0 anomalous dimensions}) --- and the OPE data of the unperturbed thermal correlator (\ref{eq:mftfirst}) to be
\be
\label{eq:quartic thermal Mellin amplitude}
M_{\beta}(s) =-{2^{-d-2} \pi^{-{d \over 2}} \over \Gamma(\Delta)\Gamma(\Delta - {d-2 \over 2})} {\zeta (2 \Delta -2 s) \Gamma \left(\Delta-\frac{d-1}{2} -s\right)\Gamma \left(2\Delta-\frac{d}{2} -s\right) \over \Gamma(\Delta+\frac{1}{2} -s) \Gamma(2 \Delta -(d-1)-s)}\, .
\ee 
The fact that the anomalous dimensions $\gamma_\ell(n)$ of~\cite{Heemskerk:2009pn} are analytic in $n$ is crucial for our construction to work. 
We further verify this result by explicitly evaluating a Witten diagram in thermal AdS. 
Note that the structure of our result~(\ref{eq:correction from bulk quartic interaction}) is reminiscent of Mellin amplitudes for $T=0$ four-point functions, hinting at a possible formulation of thermal Mellin amplitudes.

We hope that the infinite family of perturbative solutions to the thermal bootstrap constructed in this paper can be a useful playground to test ideas and make further progress. Prior to our family of solutions, very few explicit examples of thermal correlators were known. It may very well be that some of the techniques discussed in the present paper are directly applicable to a more realistic situation of CFTs with gravity duals. We comment further on this in the conclusions.

The plan of the paper is as follows. In section~\ref{sec:thermal CFT review}, we review the setup of CFTs on $S^1_\b \times \R^{d-1}$, and collect relevant details of the thermal bootstrap problem. In section~\ref{sec:thermal holography from CFT}, we present our family of solutions to the thermal bootstrap corresponding to the $T>0$ generalization of~\cite{Heemskerk:2009pn}, and show that they follow from required consistency conditions. In section~\ref{sec:thermal holography}, we verify our solution for the simplest case of a bulk quartic interaction without derivatives by computing the corresponding thermal AdS Witten diagram. In section~\ref{sec:thermal dispersion relation}, we derive a simple dispersion relation for thermal two-point functions, reconstructing it from it's discontinuity. In section~\ref{sec:uniqueness}, we discuss the uniqueness of our solutions. Finally, in section~\ref{sec:conclusion}, we conclude, and discuss interesting open directions. Appendix~\ref{app:anomalous dimensions} presents some anomalous dimensions corresponding to the setup of~\cite{Heemskerk:2009pn}.

\section{Review: CFTs on $S^1_\b \times \R^{d-1}$}
\label{sec:thermal CFT review}

We begin by reviewing salient aspects of CFTs at nonzero temperature or, equivalently, CFTs placed on the manifold $\cM_\b = S^1_\b\times \R^{d-1}$ where $\b=1/T$. Due to the scale set by the temperature --- equivalently, the size of the thermal circle --- conformal symmetry is partially broken and the set of observables are enlarged to include new thermal data. The simplest such observables are the thermal one-point functions of operators. Due to the residual symmetries, only primary even-spin symmetric traceless tensors $\cO^{\mu_1\dots \mu_J}(x)$ can have nonzero one-point functions,
\be
\<\cO^{\mu_1\dots \mu_J}(x)\>_{\cM_\b} &= \frac{b_\cO}{\b^\De}(e^{\mu_1}\cdots e^{\mu_J} -\text{traces}),
\ee where $\De$ is the dimension of $\cO$, $e^\mu$ is a unit vector along the thermal circle, and $b_\cO$ is the dynamical thermal data.

In~\cite{Iliesiu:2018fao,Iliesiu:2018zlz}, a thermal bootstrap problem was developed, where one studies two-point functions at nonzero temperature, and uses the OPE together with the KMS condition as a crossing equation to bootstrap for the thermal one-point coefficients. Consider the two-point function of identical real scalars $\phi$ at temperature $T$,
\be
\label{eq:thermal two point function}
\<\phi(x_1)\phi(x_2)\>_{\b} \equiv \<\phi(x_1)\phi(x_2)\>_{\cM_\b}
\ee where $x_1^\mu,x_2^\mu \in \cM_\b$. We set $x_2=0$ by translations, and pick coordinates $x^\mu = (\tau,\bf{x})$ where $\tau\in S_\b^1$ and ${\bf x} \in \R^{d-1}$. The KMS condition can be expressed as the periodicity of the two-point function along the thermal circle, 
\be
\label{eq:KMS condition}
\<\phi(\tau,{\bf x})\phi(0)\>_\b = \<\phi(\tau+\b,{\bf x})\phi(0)\>_\b \, .
\ee Since $\cM_\b$ is conformally flat, one can use the OPE within a radius of convergence $|x| < \b$. For the thermal two-point function, the OPE takes the form
\be
\label{eq:OPE}
\<\phi(x)\phi(0)\>_\b &= \sum_{\cO\in \phi\times \phi} \frac{a_\cO }{\b^\De} C_J^{(\nu)}\left(\frac{x\cdot e}{|x|}\right) {|x|}^{\Delta-2\Delta_\f}, \qquad \text{for } \quad |x|< \b,
\nn \\ a_\cO&\equiv f_{\f\f\cO}b_\cO \frac{J!}{2^J (\nu)_J}\, .
\ee Here, the sum runs over operators $\cO$ in the $\f\times\f$ OPE, with $\De$ and $J$ the dimension and spin of $\cO$, and $f_{\f\f\cO}$ are OPE coefficients. The $C_J^{(\nu)}$ are Gegenbauer polynomials, while $\nu =\tfrac{d-2}{2}$. It will come in handy to introduce the variables $(r,\eta)$ defined as
\be
\label{eq:define r and eta}
r \equiv |x| = \sqrt{\tau^2 + {\bf x}^2}, \qquad \eta \equiv \frac{x\cdot e}{|x|} = \frac{\tau}{\sqrt{\tau^2 + {\bf x}^2}}\, .
\ee
We refer to the kinematical factors $C_J^{(\nu)}(\eta) r^{\De-2\De_\f}$ in~(\ref{eq:OPE}) as ``thermal blocks'', and to the dynamical data $a_\cO$ as ``thermal coefficients''.

There exist $\Z$-many OPE channels, one for an expansion around each thermal image, $x^\mu \sim m \b e^\mu$ where $m\in \Z$. Each have a limited radius of convergence, $|x-\b m e|<\b$, so only adjacent OPE channels have overlapping domains of convergence. The thermal blocks are not invariant under thermal translations, and therefore the OPE~(\ref{eq:OPE}) is not manifestly KMS-invariant. The thermal bootstrap as developed in \cite{Iliesiu:2018fao,Iliesiu:2018zlz} systematically solves for thermal coefficients by imposing the consistency of overlapping OPE channels, with the KMS condition as a crossing equation. Another useful tool in this context is a thermal Lorentzian inversion formula (TLIF). Analogous to the Lorentzian inversion formula for $T=0$ four-point functions~\cite{Caron-Huot:2017vep}, the TLIF decomposes the thermal two-point function into the OPE data with manifest analyticity in the spin, $J$, of constituent operators.\footnote{For completeness, we provide a brief review of the TLIF in section~\ref{sec:inversion from dispersion}. We refer the reader to \cite{Iliesiu:2018fao,Iliesiu:2018zlz} for further details.}

Another useful set of variables that we frequently use are defined as follows. Using the residual rotation symmetry in $\R^{d-1}$, we can set $x^\mu = (\tau,|{\bf x}|,0,\dots,0)$. Defining the variables
\be
z=\tau + i |{\bf x}|,\qquad \bar z = \tau -i |{\bf x}|\, ,
\ee the thermal two-point function becomes a function of $z$ and $\bar z$. For convenience, we define
\be
g(z,\bar z)\equiv \< \phi(z,\bar z) \phi(0)\>_\b \,.
\ee Since $\b$ is the only scale in the problem, we can henceforth set $\b=1$. Now, the KMS condition, combined with the symmetry $x \rightarrow -x$,\footnote{See \cite{Iliesiu:2018fao} for the detailed discussion.} reads
\be
\label{eq:KMS in z zbar}
g(z,\bar z) = g(1-z,1-\bar z).
\ee An important and simple solution is given by the mean field theory (MFT) two-point function~(\ref{eq:mftfirst}), which in the $(z,\bar z)$ variables becomes
\be
\label{eq:MFT two-point function}
G_s(z,\bar z) = \< \phi(z,\bar z)\phi(0)\>_\b^{\text{MFT,} \ \De_\f=s} =  \sum_{m=-\infty}^\infty \frac{1}{(z+m)^s(\bar z+m)^s}\, .
\ee This solution will feature prominently in our work, so we have decided to reserve it the name $G_s$. The OPE of the MFT two-point function is particularly simple, containing only the unit operator ${\bf 1}$ and the double-trace operators $[\f\f]_{n,\ell}$ with dimension $\De = 2\De_\f + 2n+\ell$ and even spin $\ell$. Concretely, the OPE takes the form\footnote{Note that in terms of $z,\bar z$, 
\be 
\eta = \frac{1}{2}\left(\sqrt{\frac{z}{\bar z}}+\sqrt{\frac{\bar z}{z}}\right)\, . \nn
\ee }
\be
\label{eq:MFT OPE}
G_{\De_\f}(z,\bar z) = \frac{1}{(z \bar z)^{\De_\f}} + \sum_{n=0}^\infty \sum_{\ell=0,2,\dots} a_{[\f\f]_{n,\ell}} C_{\ell}^{(\nu)}(\eta) (z \bar z)^{n+\ell/2} \, ,
\ee where the MFT thermal coefficients are given by~\cite{Iliesiu:2018fao}
\be
\label{eq:MFT thermal coefficients}
a^{(0)}_{\ell}(n) \equiv a_{[\f\f]_{n,\ell}}= 2\zeta(2\De_\f+2n+\ell)\frac{(\ell+\nu)(\De_\f)_{\ell+n}(\De_\f-\nu)_{n}}{n!(\nu)_{\ell+n+1}} \, .
\ee Once again, the MFT thermal coefficients will occur frequently below, so we have reserved the name $a^{(0)}_{\ell}(n)$ for them. The observation that $a^{(0)}_{\ell}(n)$ is a meromorphic function of $n\in \C$ will be crucial for our proposal in section~\ref{sec:thermal holography from CFT}. Note the curious fact that, unlike spin $\ell$ which has a meaningful nonperturbative analytic continuation by way of lightray operators~\cite{Kravchuk:2018htv}, there is not a general known notion of analyticity in twist $n$. Nevertheless, the functions describing the MFT data are analytic functions of $n$, and we will use this property.

\section{Thermal holography from CFT}
\label{sec:thermal holography from CFT}

In this section we present the explicit form of an infinite tower of corrections  to the MFT thermal two-point function. These corrections correspond to quartic vertices in the bulk, see (\ref{eq:perturbative QFT in AdS}). At zero temperature these corrections were considered by Heemskerk, Polchinski, Penedones and Sully (HPPS)~\cite{Heemskerk:2009pn} and we will start by briefly reviewing their construction. 

The simplest nontrivial correlator to consider at zero temperature is the four-point function of identical scalar operators $\phi$ of dimension $\Delta$, denoted by ${\cal G}(u,v)$, where $u,v$ are the conformal cross-ratios $u=\frac{x_{12}^2x_{34}^2}{x_{13}^2 x_{24}^2}$, $v=\frac{x_{14}^2x_{23}^2}{x_{13}^2 x_{24}^2}$. This four-point function satisfies the crossing equation:
 \begin{equation}
v^\Delta {\cal G}(u,v) = u^\Delta {\cal G}(v,u).
 \end{equation}
At zeroth order in a holographic theory, it is given by the MFT expression 
 \begin{equation}
{\cal G}^{(0)}(u,v) = 1+u^\Delta + \left(\frac{u}{v}\right)^\Delta\, .
 \end{equation}
The intermediate operators are the identity and a tower of double-trace operators $[\phi \phi]_{n,\ell}$, of spin $\ell$ and twist $2\Delta+2n$. HPPS constructed corrections to this correlator consistent with crossing symmetry and the conformal block decomposition. It was argued that for solutions corresponding to tree-level corrections from quartic vertices in the bulk, only double-trace operators up to a certain spin acquire an anomalous dimension. If we impose that only double-trace operators up to spin $L$ acquire an anomalous dimension, then there are $(L+2)(L+4)/8$ solutions. For each correction ${\cal G}^{(1)}(u,v) $ we have a set of anomalous dimensions 
 \begin{equation}
\gamma_{0}(n), \gamma_{2}(n),\cdots,\gamma_{L}(n)
 \end{equation}
which are fixed (up to an overall constant) by crossing symmetry. In HPPS an algorithm was given to find these anomalous dimensions. It was also shown that once the anomalous dimensions are fixed, the corrections to the OPE coefficients are also fixed, and hence the full correlator:
 \begin{equation}
\gamma_{0}(n), \cdots, \gamma_{L}(n) ~\to~ {\cal G}^{(1)}(u,v)~~~\text{($T=0$ four-point function).}
 \end{equation}
We will show below that each of these solutions at $T=0$ can be extended to a solution for the thermal two-point function away from MFT:
 \begin{equation}
\gamma_{0}(n), \cdots, \gamma_{L}(n) ~\to~ g^{(1)}(z,\bar z)~~~\text{(thermal two-point function).}
 \end{equation}

\subsection{Consistency conditions}
\label{sec:consistency conditions}

The correction, $g^{(1)}(z,\bar z)$, to the MFT thermal two-point function will be fixed by the following requirements:
\begin{enumerate}
\item The KMS condition
$$g^{(1)}(z,\bar z)=g^{(1)}(m+z,m+\bar z),$$
which together with the symmetry of the problem also implies
$$g^{(1)}(z,\bar z)=g^{(1)}(1-z,1-\bar z).$$
\item Consistency with the OPE decomposition. No new operators are exchanged. The anomalous dimensions are those at $T=0$, namely $\gamma_{0}(n), \gamma_{2}(n),\cdots,\gamma_{L}(n)$.
\item Regge boundedness. In terms of variables $(r,\eta)$ defined in~(\ref{eq:define r and eta}) (with $z \bar z = r^2$ and $z+\bar z=2\eta r$), the Regge limit corresponds to taking $|\eta| \to \infty$ with fixed $r$. We impose that $g^{(1)}(z,\bar z)$ is polynomially bounded in the Regge limit. 
\item Analyticity. The function $g^{(1)}({r /w}, r w)$ is analytic in the $ w$ cut plane, see figure \ref{fig:dispersion contour}.\footnote{This analyticity was established at the nonperturbative level in \cite{Iliesiu:2018fao} using the Kaluza-Klein representation of the correlator. We will assume that it holds for the perturbative solutions of interest as well.}
\item Clustering. At large spatial separations the correlator goes to zero\footnote{Recall that due to the $\mathbb{Z}_2$ symmetry of the problem $\< \phi \>_{\beta} = 0$.}
\be
\lim_{| {\bf x}| \to \infty} g^{(1)}(\tau + i |{\bf x}| ,\tau - i |{\bf x}|) = 0 . 
\ee
\end{enumerate}
Given that the anomalous dimensions are fixed, the problem effectively reduces to finding the corrections to the thermal MFT coefficients $a^{(1)}_{n,\ell}$. Below we will write down a proposal for the full leading-order correction, $g^{(1)}(z,\bar z)$, to the MFT thermal correlator and show that it satisfies all required conditions. From this one can extract integral expressions for the OPE coefficients if desired.

\subsection{The case without derivatives}

Let's start with our proposal for the case $L=0$, corresponding to a quartic interaction without derivatives. It takes the form
\begin{equation}
\label{L0prop}
g^{(1)}_{L=0}(z,\bar z)= \oint_{\e-i \infty}^{\e+i \infty} \frac{ds}{2\pi i} \Gamma(s)^2 \Gamma(\Delta-s)^2 M_{\beta}(s) G_s(z,\bar z)
\end{equation}
where we assume $M_{\beta}(s)$ is regular at $s=0,-1,-2,\cdots$, and the contour is chosen as to pick the contribution from these poles upon deforming to the left. Provided this integral converges, it is clear that $g^{(1)}_{L=0}(z,\bar z)$ satisfies the KMS condition: The dependence on $z,\bar z$ comes through the MFT-like factor in the integrand, which satisfies the KMS condition for all $s$. The issue of convergence will be addressed later on. 

Next, we would like to show that $M_{\beta}(s)$ is fixed by the requirement that we reproduce the correct anomalous dimensions and that this proposal is consistent with the OPE decomposition. In order to do that, we will compute $g^{(1)}_{L=0}(z,\bar z)$ in a $z,\bar z$ expansion. It is convenient to organise the computation in terms of the $(r,\eta)$ variables and give the result as an expansion in powers of $r$. 

Let's first look at the contribution from $m=0$. It is clear that this is the only contribution that can produce $\log r$, and in particular it fully encodes the anomalous dimensions. Performing the sum over residues, we obtain
\begin{equation}
\left. g^{(1)}_{L=0}(z,\bar z) \right|_{m=0}= \sum_{n=0} \frac{\left(\hat M_{\beta}'(-n)-2 \hat M_{\beta}(-n) (\log r-\psi ^{(0)}(n+1))\right)}{\Gamma (n+1)^2}r^{2n}\,,
\end{equation}
where we have introduced $\hat M_{\beta}(s) = \Gamma(\Delta-s)^2 M_{\beta}(s)$. For any given $M_{\beta}(s)$, this expression is very explicit. Note that only scalar operators acquire an anomalous dimension. Furthermore, reproducing the correct anomalous dimension requires
\begin{eqnarray}
\label{eq:h in terms of gamma}
M_{\beta}(-n)&=&-\frac{\Gamma (n+1)^2 a^{(0)}_{0}(n) }{2\Gamma (n+\Delta )^2}\gamma_{0}(n)\,,
\end{eqnarray}
where $a^{(0)}_{0}(n)$ are the spin-zero thermal coefficients in MFT, presented in~(\ref{eq:MFT thermal coefficients}), and $\gamma_{0}(n)$ the corresponding anomalous dimension, given in~(\ref{eq:spin 0 anomalous dimensions}). As discussed in \cite{Heemskerk:2009pn}, the anomalous dimensions $\gamma_{0}(n)$ of quartic vertices can be analytically continued in $n$. The MFT thermal coefficients admit a natural analytic continuation in $n$ given by their explicit expression in~(\ref{eq:MFT thermal coefficients}) for $n\in \C$. We propose that $M_{\beta}(s)$ is given by the above relation, where we take $s=-n\in \C$ with the aforementioned analytic expressions for $\gamma_{0}(n)$ and $a_{0}^{(0)}(n)$. Now we turn to the contribution from the $m \neq 0$ terms. Expanding in powers of $r$ and then performing the sum we obtain
\begin{eqnarray}
 \sum_{m \neq 0}\frac{1}{(m-z)^s(m-\bar z)^s} &=& 2 \zeta (2 s)+ 2 s \left(2 \eta ^2 (s+1)-1\right) \zeta (2 (s+1))r^2 + \cdots.
\end{eqnarray}
This expansion contains only even powers of $r$ and $\eta$, and the power $r^{2n}$ is multiplied by a polynomial in $\eta^2$ of degree $n$. Plugging this expansion into the integrals, it remains to discuss the issue of convergence. The integrals to be performed are of the form
\begin{equation}
 \int_{\e-i \infty}^{\e+i \infty} \frac{ds}{2\pi i} \Gamma(s)^2 \Gamma(\Delta-s)^2 \zeta (2 s) \zeta (2 (\Delta-s))
\end{equation}
up to rational functions, which are not important for this discussion. This is a convergent integral, since the factor $\Gamma(s)^2 \Gamma(\Delta-s)^2$ decays exponentially along both directions for $s= \epsilon+ i \mu$, with $\mu$ real. For any fixed $d,\Delta$, and any term in the small $r$ expansion, the resulting integrals can be evaluated numerically, to any desired precision.  Summarising, the integral has the following expansion:
\begin{equation}
g^{(1)}_{L=0}(z,\bar z)  = \left. g^{(1)}_{L=0}(z,\bar z) \right|_{m=0} +  P_0(\eta^2) + P_2(\eta^2) r^2 +P_4(\eta^2) r^4+\cdots,
\end{equation}
where for a given $d,\Delta$ the polynomials $P_{n}(\eta^2)$ can be computed to any desired accuracy. 

Let's now discuss the OPE expansion of $g^{(1)}_{L=0}(z,\bar z)$. For the total answer (thermal MFT plus corrections) it should take the form 
\begin{equation}
g^{(0)}(z,\bar z) +\lambda g^{(1)}_{L=0}(z,\bar z) +\cdots = \sum_{n,J} a_{n,J} C_{J}^{(\nu)}(\eta) r^{2n+J+\lambda \gamma_{J}(n)+\cdots}
\end{equation}
As already mentioned, an anomalous dimension leads to a term proportional to $\log r$, and $g^{(1)}_{L=0}(z,\bar z)$ produces an anomalous dimension for operators with $J=0$ and $n=0,1,\cdots$, as expected. Corrections to the OPE coefficients take the form
\begin{equation}
\sum_{n,J} a^{(1)}_{n,J} C_{J}^{(\nu)}(\eta) r^{2n+J}\,.
\end{equation}
Consistency with the OPE implies that our answer should be expandable as above, with $J=0,2,\cdots$ and $n=0,1,2,\cdots$. The Gegenbauer polynomials of even spin are given by even polynomials of $\eta$ of degree $J$. The expansion of $g^{(1)}_{L=0}(z,\bar z)$ around small $z,\bar z$ is exactly of this form, and hence consistent with the OPE. 

Before proceeding to the general case, note that the proposed integral converges for any (non-integer) value of $z,\bar z$. This is due to the strong exponential decay of the factor $\Gamma(s)^2 \Gamma(\Delta-s)^2$. Finally, let's discuss the Regge limit and the analyticity properties of our proposal. One can show that in the Regge limit the main contribution arises from the $m=0$ term,
\begin{equation}
\sum_{m=-\infty}^\infty \frac{1}{(m-z)^s(m-\bar z)^s} = \frac{1}{r^{2s}} + {\cal O}\left(\frac{1}{r^s \eta^s}\right),
\end{equation}
where we assume ${\rm Re}[s]>0$. Plugging this into the integral we find that in the Regge limit
\begin{equation}
g^{(1)}_{L=0}(z,\bar z) \sim f(r)
\end{equation}
and is indeed polynomially bounded. Note also that since the integral is convergent as long as the MFT correlator is finite, the integral is analytic in the same domain as $G_s(z,\bar z)$ in the complex $z,\bar z$-planes, as the convergent integral cannot produce new poles or cuts. This implies, in particular, that our proposal satisfies the required analyticity property in the complex $w$-plane, since the MFT correlator does so.

\subsection{General proposal}
In the following, we propose the general answer for the thermal two-point functions corresponding to quartic interactions with derivatives. This corresponds, at zero temperature, to the solutions described in \cite{Heemskerk:2009pn} for a maximum spin $L > 0$. For solutions truncated at higher values of the spin we propose an expression similar to (\ref{L0prop}), but with derivatives $\partial_z, \partial_{\bar z}$ acting on the MFT-like factor:
\begin{equation}
\label{Lprop}
g^{(1)}_{L}(z,\bar z) = \int_{\e-i \infty}^{\e+i \infty}  \frac{ds}{2\pi i} \Gamma(s)^2 \Gamma(\Delta-s)^2 \mathfrak{M}^{L}(s,\partial_z,\partial_{\bar z}) G_s(z,\bar z)\, .
\end{equation}
Note that in order for the derivatives to preserve the $z \leftrightarrow \bar z$ symmetry, as well as the KMS condition, the differential operator should be invariant under the following transformations
\begin{equation}
\mathfrak{M}^{L}(s,\partial_z,\partial_{\bar z})=\mathfrak{M}^{L}(s,\partial_{\bar z},\partial_z),~~~\mathfrak{M}^{L}(s,\partial_z,\partial_{\bar z})=\mathfrak{M}^{L}(s,-\partial_z,-\partial_{\bar z}) \, .
\end{equation}
Furthermore, note that the action of $\partial_z \partial_{\bar z}$ can be absorbed into the $s$ dependence, since 
\begin{equation}
\partial_z \partial_{\bar z} \frac{1}{(m-z)^s(m-\bar z)^s} = \frac{s^2}{(m-z)^{s+1}(m-\bar z)^{s+1}} \, ,
\end{equation}
and we can go back to the original form after shifting $s \to s-1$. On the other hand, the action of a term like $\partial_z^L+\partial_{\bar z}^L$ would produce an anomalous dimension for spin $L$ operators (and lower). Indeed, pulling out this operator outside the integral, and acting on $r^{2n}$ we obtain
\begin{eqnarray}
\left( \partial_z^2+\partial_{\bar z}^2 \right) r^{2n} &=& \left( \partial_z^2+\partial_{\bar z}^2 \right) z^{n} {\bar z}^n =2 \left(2 \eta ^2-1\right) (n-1) n r^{2n-2} \\
\left( \partial_z^4+\partial_{\bar z}^4 \right) r^{2n} &=& 2 \left(8 \eta ^4-8 \eta ^2+1\right) (n-3) (n-2) (n-1) n r^{2n-4}
\end{eqnarray}
and so on, generating the correct powers of $\eta$ and $r$ for an operator of higher (even) spin --- $J=2,0$ for the first line, and $J=4,2,0$ for the second line --- and the correct twist. This leads to the proposal (\ref{Lprop}) where the operator $\mathfrak{M}^L(s,\partial_z,\partial_{\bar z})$ is given by
\begin{equation}
\mathfrak{M}^{L}(s,\partial_z,\partial_{\bar z})= M_{\beta,0}(s) + M_{\beta,2}(s) \left( \partial_z^2+\partial_{\bar z}^2 \right)  + \cdots+ M_{\beta,L}(s) \left( \partial_z^L+\partial_{\bar z}^L \right) \, .
\end{equation}
The operator $\mathfrak{M}^{L}(s,\partial_z,\partial_{\bar z})$ has the correct symmetries such that $g^{(1)}_{L}(z,\bar z)$ satisfies the KMS condition. The small $r$ expansion of $g^{(1)}_{L}(z,\bar z)$ can be performed in exactly the same way as for $g^{(1)}_{L=0}(z,\bar z)$. We can actually pull out all derivative terms, compute the expansions, and then act with them. From this, and our discussion for $L=0$, it is clear that the functions $M_{\beta,0}(s),M_{\beta,2}(s),\cdots, M_{\beta,L}(s)$ are again fixed by requiring the correct anomalous dimensions for operators with spin $\ell=0,2,\cdots,L$. Furthermore, note that the action of $\partial_z^L+\partial_{\bar z}^L$  will not spoil consistency with the OPE. Finally, we can also study this solution in the Regge limit. Again the leading contribution arises from the $m=0$ term and we find 
\begin{equation}
g^{(1)}_{L}(z,\bar z) \sim \eta^L f(r).
\end{equation}
As with the solution with $L=0$, the integral is once again convergent as long as the MFT correlator and its derivatives $(\partial_z^L+\partial_{\bar z}^L)G_s(z,\bar z)$ are finite for given $z,\bar z$. Since derivatives cannot produce new poles or cuts at previously analytic points of a meromorphic function, the solution is analytic in the same domain as the MFT correlator. Thus, our general solution satisfies the analyticity condition in the $w$-plane.

\subsection{Example: maximal spin two}
As an example, let us write down our proposed answer for the case $L=2$, namely $g^{(1)}_{L=2}(z,\bar z) $, in terms of the anomalous dimensions of double-trace operators with spin zero and two --- $\gamma_{0}(n)$ and $\gamma_{2}(n)$. We focus on the part of the answer proportional to $\log r$. We obtain
\begin{equation}
g^{(1)}_{L=2}(z,\bar z)  = -2 \log r \sum_{n=0} \frac{\Gamma(\Delta+n)^2}{\Gamma(n+1)^2}\left(M_{\beta,0}(-n) + M_{\beta,2}(-n) \left( \partial_z^2+\partial_{\bar z}^2 \right)  \right)r^{2n} + \cdots \, .
\end{equation}
Acting with the derivative operators and shifting the variable $n$ we obtain
\begin{equation}
\left. g^{(1)}_{L=2}(z,\bar z)  \right|_{\log r} = -2 \sum_{n=0} \frac{\Gamma(\Delta+n)^2}{\Gamma(n+1)^2}\left(M_{\beta,0}(-n) + \frac{2 \left(2 \eta ^2-1\right) n  (\Delta +n)^2}{n+1} M_{\beta,2}(-n-1)\right)r^{2n} .
\end{equation}
This should be equated to the corresponding contribution to the OPE from the anomalous dimension
\begin{eqnarray}
\left. g^{(1)}_{L=2}(z,\bar z)  \right|_{\log r} &=& \sum_{n=0} \left( a^{(0)}_{0}(n) \gamma_0(n) C_0^{(\nu)}(\eta) r^{2n} + a^{(0)}_{2}(n) \gamma_2(n) C_2^{(\nu)}(\eta) r^{2n+2}  \right) \\
&=& \sum_{n=0}  \left( a^{(0)}_{0}(n) \gamma_0(n) + a^{(0)}_{2}(n-1) \gamma_2(n-1) \frac{d \eta^2-1}{d-1} \right)r^{2n} \, ,
\end{eqnarray}
where we have used $C_0^{(\nu)}(\eta)=1,C_2^{(\nu)}(\eta)=\frac{d \eta^2-1}{d-1}$ and we take $a^{(0)}_{2}(-1)=0$. Equating different powers of $\eta$, we find
\begin{eqnarray}
M_{\beta,0}(-n)&=&-\frac{\Gamma (n+1)^2 (2 (d-1) a^{(0)}_{0}(n) \gamma_0(n)+(d-2) a^{(0)}_{2}(n-1)\gamma_2(n-1))}{4 (d-1) \Gamma (n+\Delta )^2} \, , \\
M_{\beta,2}(-n) &=& -\frac{d n a^{(0)}_{2}(n-2) \gamma_2(n-2) \Gamma (n)^2}{8 (d-1) (n-1) (\Delta +n-1)^2 \Gamma (n+\Delta -1)^2} \, .
\end{eqnarray}
Let us compute this for a precise example. Consider for instance $d=4$ and $\Delta=2$ and the anomalous dimensions corresponding to a quartic vertex $(\nabla \Phi)^4$. In this case, we obtain
\begin{eqnarray}
M_{\beta,0}(s)&=& \frac{(s-5) \left(20 s^6-360 s^5+2613 s^4-9752 s^3+21996 s^2-34272 s+24192\right) \zeta (8-2s)}{72 (2 s-9) (2 s-7) (2 s-5) (2 s-3)}\, , \nonumber\\
M_{\beta,2}(s) &=& \frac{(s-5) (s-4) (s-3) s (s+1) \zeta(6-2 s)}{144 (2 s-7) (2 s-5) (2 s-3) (2 s-1)} \, .
\end{eqnarray}
In particular, note that $M_{\b,\ell}(s)$ do not have poles at $s=0,-1,\cdots$, in agreement with our assumptions. 

\section{Holographic solution from bulk interaction}
\label{sec:thermal holography}

We are interested in computing holographic thermal two-point functions. The geometry of the thermal setup we are interested in has $S^1_\b\times \R^{d-1}$ on the boundary. In the bulk we consider thermal AdS, which is the correct setting for QFTs in AdS at finite temperature. In the case where the theory in AdS is gravitational, the relevant bulk geometry is the one of black brane \cite{Witten:1998zw}. We defer this very interesting and harder case to future work, and proceed with a computation in thermal AdS.

\subsection{Thermal AdS review}
\label{subsec:thermal AdS}

Thermal AdS is a quotient of Euclidean AdS (EAdS) by $\Z$, so many properties can be inferred from EAdS. EAdS$_{d+1}$ itself is given by the hyperboloid
\be
P\cdot P \equiv -P_0^2 + P_1^2+\dots +P_d^2 + P_{d+1}^2 =-R^2
\ee
in embedding space $\R^{1,d+1}$. Defining lightlike embedding space coordinates
\be
P^A = (P_-,P_+,P_i)
\ee
with $P_\pm = P_0\pm P_{d+1}$, we can choose Poincar\'e patch coordinates 
\be
\label{eq:poincare coordinates}
P^A = (R^2/z, (z^2 +x^2)/z, R \, x^\mu/z),
\ee with $x^\mu=(\tau,{\bf x})$. 
In these coordinates, the EAdS metric is given by
\be
\label{eq:AdS metric}
ds^2 &= R^2 \frac{1}{z^2} (dz^2 + d\tau^2 + d{\bf x}^2),
\ee 
with thermal AdS given by the space with the identification $\tau \sim \tau +\b$. The AdS length is $R$, and to match with the boundary CFT we set $R=1$ henceforth as usual, and once again set $\b=1$. We are interested in the case $d\ge 2$, in which case AdS$_{d+1}$ has a connected boundary. The boundary is situated at $z=0$, obtained via the usual limit
\be
X^A = \lim_{z\rightarrow 0} z P^A = (1,\tau^2 + {\bf x}^2, \tau,{\bf x}).
\ee Note the identification of $\tau \sim \tau +\b$ in the bulk indeed produces the expected boundary $\cM_\b$. We will refer to thermal AdS with temperature $T=1/\b$ as $\cB_\b$. 

The AdS/CFT correspondence implies that free fields propagating in thermal AdS, $\cB_\b$, are dual to MFT in the boundary $\cM_\b$. This can be verified by studying the free field propagator in $\cB_\b$ and matching it to the MFT thermal two-point function in the boundary limit. Consider a real scalar field $\phi(X)$ of dimension $\De$ on the boundary, and it's bulk dual $\Phi(P)$. The bulk-to-bulk propagator in AdS is given by
\be
\label{eq:AdS bulk propagator}
\< \Phi(P)\Phi(Q)\>^{(0)} = C_\De u^{-\De}\, {}_2F_1(\De, \De-\frac{d}{2}+\frac{1}{2},2\De-d+1;-\frac{4}{u}),
\ee where 
$u = (P-Q)^2 = -2-2P\cdot Q$, and the natural normalization of the bulk fields gives the factor 
\be
C_\De = \frac{\G(\De)}{2\pi^{d/2}\G(\De-d/2+1)}\, .
\ee
The propagator in thermal AdS is constructed from the AdS propagator via the method of images,\footnote{A more general way to construct the propagator is to sum over the images of both points, and divide by the volume of the quotient group, $\vol \Z$. The infinite sum over simultaneous thermal translations of both points can then be gauge fixed, canceling the infinite volume factor, and one is left with a sum over relative translations of the points. This point of view is helpful when considering more general diagrams than we do here.
}
\be
\label{eq:thermal AdS bulk propagator}
\< \Phi(P)\Phi(Q)\>_\b^{(0)} \equiv \sum_{m=-\infty}^\infty \< \Phi(P_{m})\Phi(Q)\>^{(0)}\,,
\ee where $P_{m}$ denotes the image of a point $P$ after $m$ thermal translations $\tau\rightarrow \tau + m$ (recall that we have set $\b=1$). We define the boundary field $\phi$ by the limit
\be
\label{eq:boundary field definition}
\phi(X) = C_\Delta^{-1/2} \lim_{z\rightarrow 0} z^{-\De} \Phi(P)\,.
\ee The extra factor of $C_\Delta^{-1/2}$ is so that the boundary operators have the standard CFT normalization without the factor of $C_\De$ in their vacuum two-point function. 
Taking one of the fields in~(\ref{eq:thermal AdS bulk propagator}) to the boundary, we obtain the bulk-to-boundary propagator,
\be
\label{eq:thermal AdS bulk to boundary propagator}
\< \phi(X)\Phi(Q)\>_\b^{(0)} \equiv \sum_{m=-\infty}^\infty \< \phi(X_{m})\Phi(Q)\>^{(0)}\,,
\ee where
\be
\label{eq:AdS bulk to boundary propagator}
\< \phi(X)\Phi(Q)\>^{(0)} = \frac{C_\De^{1/2}}{(-2 X\cdot Q)^{\De}}
\ee is the AdS bulk-to-boundary propagator. Note that by taking both points to the boundary we indeed recover the MFT two-point function,
\be
\label{eq:boundary MFT}
\< \phi(X)\phi(Y)\>_\b^{(0)} = \sum_{m=-\infty}^\infty \frac{1}{(-2X_m\cdot Y)^\De}\, .
\ee Note that we indeed have the standard CFT normalization thanks to our choice in~(\ref{eq:boundary field definition}).

\subsection{Contribution of the $\l_0 \Phi^4$ contact diagram}
\label{subsec:lambda phi^4 diagram}

\begin{figure}[tb]
\begin{center}
\begin{tikzpicture}[xscale=1,yscale=1]
\draw[thick] (0,0) circle (2cm);
\filldraw[gray] (0,0) circle (.2cm);
\draw[blue] (0,0) circle (1cm);
\draw[blue] (-1.78201,.9079) -- (-1,0);
\draw[blue] (-1.78201,-.9079) -- (-1,0);
\node[left] at (-1.78,1) {$\f(x_1)$};
\node[left] at (-1.78,-1) {$\f(x_2)$};
\filldraw[red] (-1,0) circle (.05cm);
\end{tikzpicture}
\caption{The thermal AdS Witten diagram for the leading order correction to the $\<\f(x_1)\f(x_2)\>_\b$ thermal correlator from a bulk $\l_0 \Phi^4$ interaction. The interaction point (red) is integrated over thermal AdS. The gray circle in the center is meant to remind us that there is a periodic direction in the bulk, and propagators (blue) can wind around that direction. In computing the diagram, one sums over these thermal images of the endpoints of propagators.}
\label{fig:thermal Witten diagram}
\end{center}
\end{figure}
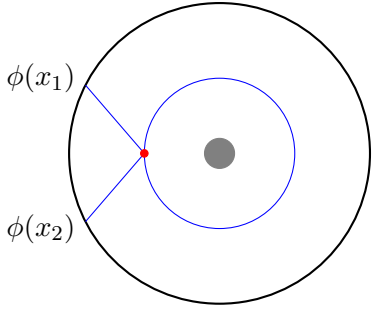

Now, let us consider a bulk $\l_0 \Phi^4$ interaction in (\ref{eq:perturbative QFT in AdS}). We will compute the leading correction to the boundary thermal two-point function~(\ref{eq:boundary MFT}) and see that we recover the answer given by our general result in section~\ref{sec:thermal holography from CFT}. The correction to the thermal two-point function is given by the thermal AdS Witten diagram presented in figure~\ref{fig:thermal Witten diagram}, which is evaluated by the integral
\be
\label{eq:thermal Witten diagram}
\<\phi(X_1)\phi(X_2)\>_\b\big|_{\lambda_0^1} =- \frac{1}{2}\int_{\cB_\b} d^{d+1} P \sqrt{g} \< \phi(X_1)\Phi(P)\>_\b^{(0)} \< \Phi(P)\Phi(P)\>_\b^{(0)} \< \Phi(P)\phi(X_2)\>_\b^{(0)}\, ,
\ee 
where ${1 \over 2}$ is the symmetry factor of the diagram. Let us point out that the bulk-to-bulk propagator is between coincident points. This naively results in a divergence, but is not immediately meaningless, as there are finite contributions from the images. The divergent term coming from the summand with image number $m=0$ in the propagator~(\ref{eq:thermal AdS bulk propagator}) as $Q\rightarrow P$  is the usual mass renormalization which we set to zero. 
Therefore, writing out the propagators, the diagram is given by 
\be
\< \f(X_1)\f(X_2)\>_\beta \big|_{\lambda_0^1}  &=- \frac{1}{2}\int_{\cB_\b}  d^{d+1}P \sqrt{g} \sum_{m,n} \sum_{p\ne 0} \frac{C_\De^2}{(-2X_{1,m}\cdot P)^\De (-2X_{2,n}\cdot P)^\De} \nn
\\ &\qquad \qquad \times u_p^{-\De}\, {}_2F_1(\De, \De-\frac{d}{2}+\frac{1}{2},2\De-d+1;-4/u_p),
\ee where
\be
u_p = (P_p-P)^2\,. 
\ee 

In order to integrate the hypergeometric function, we write it in a Barnes representation,
\be
\label{eq:Barnes rep}
_2F_1(a,b,c; -x)&= \frac{\G(c)}{\G(a)\G(b)} \int_{s_0-i\infty}^{s_0+i\infty} \frac{ds}{2\pi i} \frac{\G(s)\G(a-s)\G(b-s)}{\G(c-s)} x^{-s} \, ,
\ee where the contour sits at real part $0<s_0 <\min(\mathrm{Re} \ a,\mathrm{Re} \ b)$. 
For us, this means placing the contour at $0<s_0<\Delta-(d-1)/2$. The next step is to swap the contour integral with the bulk integral and the sum over images. Pulling out the contour integral, we have 
\be
\label{eq:thermal Witten Barnes representation}
&\< \f(X_1)\f(X_2)\>_\beta \big|_{\lambda_0^1}  \nn
\\ &=- \int_{s_0-i\infty}^{s_0+i\infty} \frac{ds}{2\pi i} \frac{\Gamma (2 \Delta -d+1) C_\De^2}{\Gamma (\Delta ) \Gamma \left(\Delta-\frac{d}{2} +\frac{1}{2}\right)}\frac{2^{-2s-1} \Gamma (s) \Gamma (\Delta -s) \Gamma \left(\Delta -\frac{d}{2}+\frac{1}{2}-s\right) }{ \Gamma (2 \Delta-d +1-s)} f_{s-\De,\De}(X_1,X_2),
\ee
where we have defined the bulk master integral
\be
f_{a,\De}(X_1,X_2)\equiv \int_{\cB_\b}  d^{d+1}P \sqrt{g} \sum_{m,n} \sum_{p\ne 0} \frac{1}{(-2X_{1,m}\cdot P)^\De (-2X_{2,n}\cdot P)^\De} u_p^{a} \, .
\ee

To compute the bulk integral, we go to Poincar\'e patch coordinates~(\ref{eq:poincare coordinates}), and matching boundary coordinates
\be
X_i^A=(1,\tau_i^2+{\bf x}_i^2,\tau_i,{\bf x}_i),
\ee with $X_{i,m}$ denoting the boundary point obtained by a thermal translation $\tau_i\rightarrow \tau_i+m$, as before. With these coordinates, we have
\be
u_p = \frac{2 p^2}{z^2}\, ,
\ee and
\be
-2 X_i\cdot P = \frac{(\tau-\tau_i)^2+({\bf x}-{\bf x}_i)^2+z^2}{ z}\, .
\ee The volume form is $\sqrt{g}=z^{-d-1}$. Furthermore, we borrow a standard technique from Witten diagram computations, and write
\be
\frac{1}{(-2 X_i\cdot P)^\De} &= z^\De \frac{1}{\Gamma(\De)} \int_0^\infty \frac{ds_i}{s_i} s_i^{\De} e^{-s_i ((\tau-\tau_i)^2+({\bf x}-{\bf x}_i)^2+z^2)}\, .
\ee Putting the pieces together, we can perform the sum over $p$, leaving us with the integrals
\be
f_{a,\De}(X_1,X_2) &=\frac{ 2\zeta(-2a) }{\G^2(\De)} \int_0^\infty ds_1 s_1^{\De-1} \int_0^\infty ds_2 s_2^{\De-1} \int_0^1 d\tau \sum_{m,n=-\infty}^\infty e^{-s_1(\tau-\tau_1-m)^2-s_2(\tau-\tau_2-n)^2} 
\nn \\ &\qquad \times \int_{\R^{d-1}} d^{d-1}{\bf x}\, e^{-s_1({\bf x}-{\bf x}_1)^2-s_2({\bf x}-{\bf x}_2)^2} \int_0^\infty dz z^{2\De-2a-d-1}
 e^{-(s_1+s_2)z^2} \, .
\ee
The ${\bf x}$ and $z$ integrals are straightforward to evaluate. The $\tau$ integral and the sums over images require more care, but can be evaluated as follows. Summing over the images first yields elliptic $\vartheta$ functions, which can be simplified by $SL(2,\Z)$ modular $s$-transforms:
\be
\sum_{m,n=-\infty}^\infty e^{-s_1(\tau-\tau_1-m)^2-s_2(\tau-\tau_2-n)^2} &= \vartheta_3\left(  \tau_1-\tau; \frac{i \pi }{s_1}\right) \vartheta_3\left(\tau_2-\tau; \frac{i \pi }{s_2}\right)\, .
\ee The $\vartheta_3$ function is defined as
\be
\vartheta_3(z;\tau) = \sum_{m=-\infty}^\infty e^{2\pi i z m} e^{i\pi \tau m^2},
\ee and satisfies the modular $s$-transformation
\be
\vartheta_3\left(\frac{z}{\tau};-\frac{1}{\tau}\right) = \sqrt{- i \tau}e^{\frac{i \pi z^2}{\tau}} \vartheta_3(z,\tau).
\ee The integral over $\tau$ can now be evaluated, and it gives a nice pairing on $\vartheta$ functions,
\be
\int_0^1 d\tau \vartheta_3\left(  \tau_1-\tau; \frac{i \pi }{s_1}\right) \vartheta_3\left(\tau_2-\tau; \frac{i \pi }{s_2}\right) &= \vartheta_3 \left(\tau_1-\tau_2;i \pi \frac{ (s_1+s_2)}{s_1 s_2} \right) \, .
\ee With a final $s$-transform, we can write 
\be
f_{a,\De}(X_1,X_2) &= \frac{ \zeta(-2a) \pi^{d/2} }{\G^2(\De)\Gamma \left(\De -\frac{d}{2}-a \right)} \int_0^\infty ds_1 s_1^{\De-1} \int_0^\infty ds_2 s_2^{\De-1}  \nn \\ &\qquad \times \sum_{m=-\infty}^\infty e^{-\tfrac{s_1 s_2}{s_1+s_2}((\tau_1-\tau_2+m)^2 +({\bf x}_1- {\bf x}_2)^2)} \left(s_1+s_2\right)^{a-\De} \, .
\ee Note that the result is a function of the thermal cross ratios $z$ and $\bar z$, as it should be, since $(\tau_1-\tau_2+m)^2 +({\bf x}_1- {\bf x}_2)^2 = (z+m)(\bar z+m)$. Finally, we can evaluate the remaining $s_i$ integrals by redefining $s_2 \rightarrow s_1 s_2$, obtaining
\be
f_{a,\De}(X_1,X_2) &= \frac{\pi ^{d/2} \zeta(-2a) \Gamma (-a)^2 \Gamma (a+\Delta ) \Gamma \left(-a+\Delta -\frac{d}{2}\right)}{ \Gamma (-2 a) \Gamma (\Delta )^2} G_{a+\De}(z,\bar z) \, .
\ee

Returning to the computation of the thermal Witten diagram~(\ref{eq:thermal Witten Barnes representation}) and plugging in $f_{s-\De,\De}$, we obtain our final answer
\be
\label{eq:lambda Phi^4 final answer}
\< \f(X_1)\f(X_2)\>_\beta \big|_{\lambda_0^1}   &=  \int_{s_0-i\infty}^{s_0+i\infty} \frac{ds}{2\pi i} \Gamma (s)^2 \Gamma (\Delta -s)^2 M_{\beta}(s) \, G_s(z,\bar{z})
\ee where
\be
M_{\beta}(s) &=-{2^{-d-2} \pi^{-{d \over 2}} \over \Gamma(\Delta)\Gamma(\Delta - {d-2 \over 2})} {\zeta (2 \Delta -2 s) \Gamma \left(\Delta-\frac{d-1}{2} -s\right)\Gamma \left(2\Delta-\frac{d}{2} -s\right) \over \Gamma(\Delta+\frac{1}{2} -s) \Gamma(2 \Delta -(d-1)-s)} .
\ee 
Recall that the contour is placed at $0<s_0<\Delta-(d-1)/2$, and therefore upon enclosing the contour to the left, the integral picks out the desired poles at $s=0,-1,\cdots$. Note that the $\zeta$ function produced a new pole at $s=\De-1/2$, which safely sits to the right of our contour.\footnote{Since we have $\De>1/2$ --- in order for MFT to make sense --- and $d\ge 2$.} Finally, note that $M_{\beta}(s)$ is identical to the result quoted in~(\ref{eq:quartic thermal Mellin amplitude}), and indeed takes the expected form proposed in~(\ref{eq:h in terms of gamma}),
\be
M_{\beta}(s) = -\frac{\G(-s+1)^2}{2\G(-s+\De)^2} a_{0}^{(0)}(-s) \g_{0}(-s) \, ,
\ee explicitly agreeing with our proposal and providing a nontrivial check.

\section{Thermal dispersion relations}
\label{sec:thermal dispersion relation}

\begin{figure}[tb]
\begin{center}
\includegraphics[width=.8\textwidth]{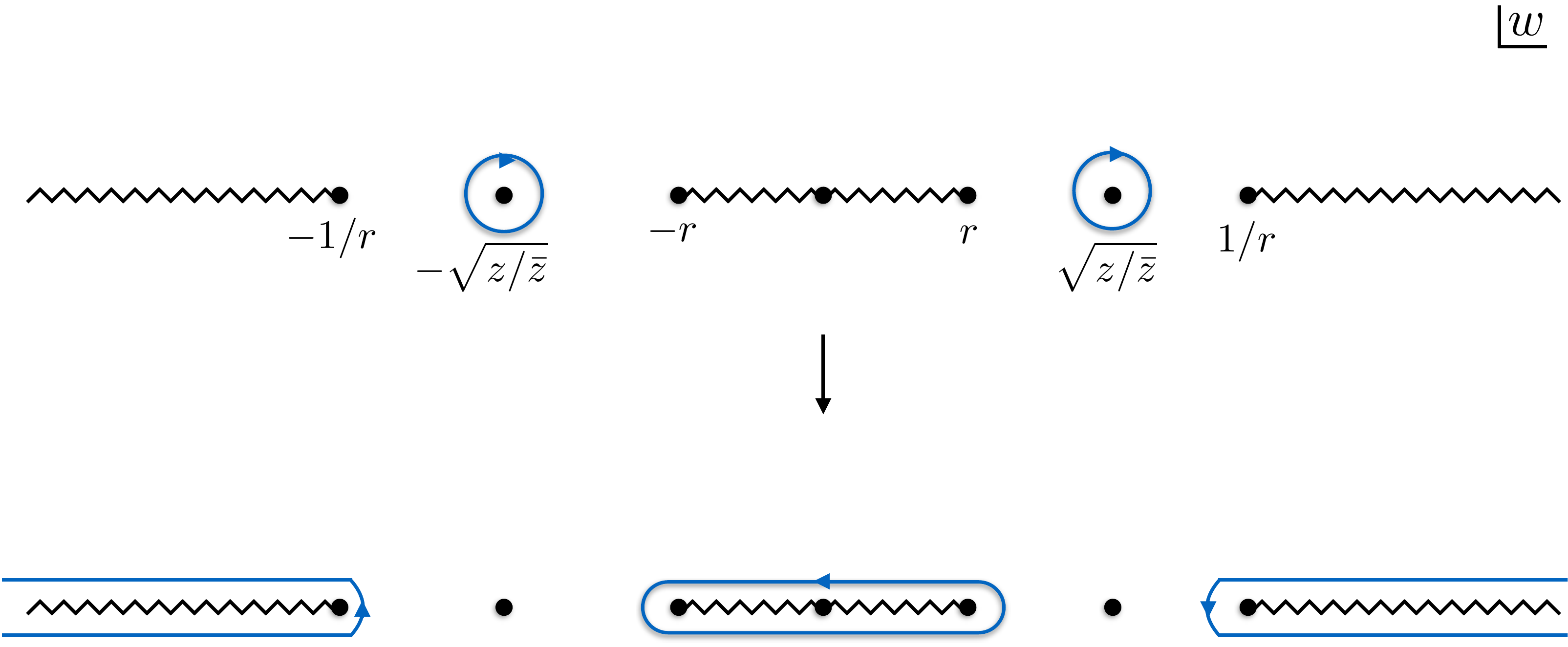}
\caption{Analytic structure of the integrand of~(\ref{eq:dispersion starting relation}) in the complex $w$-plane. The thermal two-point function $g(r/w,r w)$ has OPE cuts starting at $\pm r$ and $\pm 1/r$, where $r = \sqrt{z \bar z}$. The kernel $K(z,\bar z; w)$ has poles at $ w=0, \pm \sqrt{z/\bar z}$. The initial contour $\cC_0$ wraps the poles at $ w = \pm \sqrt{z/\bar z}$ as pictured above. Deforming the contour, one obtains the contour along the cuts as depicted below the arrow, yielding an integral over the discontinuities of $g$.}
\label{fig:dispersion contour}
\end{center}
\end{figure}

Let us now turn to deriving a thermal dispersion relation. The goal is to reconstruct $g(z,\bar z)$ with given analytical properties from its discontinuity $\Disc[g(z,\bar z)]$ (to be defined precisely below).
For this purpose, it is useful to introduce the following kernel:
\be
K (z,\bar z;  w ) = {1 \over  w} {1 \over 1 -  w^2 {\bar z \over z}} . 
\ee
The kernel above has simple poles at $ w = \pm {\sqrt{z} \over \sqrt{\bar z}}$ with residue $- {1 \over 2}$ each. We can use this fact to write the following identity:
\be
\label{eq:dispersion starting relation}
g(z,\bar z)  = \oint_{{\cal C}_0} {d \bar w \over 2 \pi i} K (z,\bar z;  w) g \left( {\sqrt{z \bar z} \over w} , \sqrt{z \bar{z} }\  w \right),
\ee
where ${\cal C}_0$ can be found in figure~\ref{fig:dispersion contour}. It wraps clockwise the poles $w = \pm {\sqrt{z} \over \sqrt{\bar z}}$ of the kernel $K (z,\bar z;w )$, and we used the fact that $g(z, \bar z) = g(\bar z, z) = g(-z, - \bar z)$. 

We now deform the contour, wrapping it around the cuts instead. Using the fact that 
\be
\lim_{| w| \to \infty} K (z,\bar z;  w ) \sim {1 \over w^3}
\ee
we can drop the contributions of the arcs at infinity for $| g \left( {\sqrt{z \bar z} / w} , \sqrt{z \bar{z} }\  w \right) | < |w|^2$ which we assume to be the case.
 
There are three cuts in figure ~\ref{fig:dispersion contour}. To relate them to each other it is useful to note the following useful property of the kernel:
\be
\label{eq:identity}
-{1 \over w^2} K (z,\bar z; {1 \over w}) = K (\bar z, z; w ) - {1 \over  w} ,
\ee
where note that $z$ and $\bar z$ are exchanged in the RHS of (\ref{eq:identity}). In this way we get the following result for the contribution of the cuts
\be
\label{eq:general thermal dispersion relation}
g(z,\bar z) &= {\rm Res}_{ w = 0} {1 \over  w} \left( {1 \over 1 -  w^2 {\bar z \over z}} + {1 \over 1 - w^2 {z \over \bar z}} - 1 \right) g \left( {\sqrt{z \bar z} \over w} , \sqrt{z \bar{z} }\  w \right)  \nn \\
&\quad+ {1 \over \pi} \int_{0}^{\sqrt{z \bar z}} {d w \over w} \left( {1 \over 1 -  w^2 {\bar z \over z}} + {1 \over 1 - w^2 {z \over \bar z}} - 1 \right)  \Disc[g\left( {\sqrt{z \bar z} \over w} , \sqrt{z \bar{z} }\  w \right) ] ,
\ee
where
\be
\Disc[g(z,\bar z)] \equiv \frac{1}{i} (g(z+i\e,\bar z) - g(z-i\e,\bar z)) .
\ee
Convergence of the integral in the second line of (\ref{eq:general thermal dispersion relation}) requires that $ \Disc[g\left( {\sqrt{z \bar z} / w} , \sqrt{z \bar{z} }\  w \right) ] $ goes to zero in the Regge limit $|w| \to 0$. Otherwise, we should consider subtractions, which we will discuss below.

Coordinate space dispersion relations were recently also discussed in the context of the vacuum four-point function in \cite{Bissi:2019kkx,Carmi:2019cub}.

\subsection{Dispersion relations in MFT}
\label{subsec:disprelMFT}

Let us see how the dispersion relations above work in MFT. To this extent we can rewrite
\be
\< \phi(z,\bar z)\phi(0)\>_\b^{\text{MFT}} = \frac{1}{z^{\Delta_\phi} \bar z^{\Delta_\phi}}  +  \sum_{m=1}^\infty \left( \frac{1}{(z+m)^{\Delta_\phi}(\bar z+m)^{\Delta_\phi}} + \frac{1}{(z-m)^{\Delta_\phi}(\bar z-m)^{\Delta_\phi}} \right) ,
\ee
where we separated the $m=0$ contribution and combined the $\pm m \neq 0$ terms. The $m=0$ term ${z^{-\Delta_\phi} \bar z^{-\Delta_\phi}}$ is reconstructed from the first line in the dispersion relations (\ref{eq:general thermal dispersion relation}), namely as a residue at $w=0$. For the $m \neq 0$ terms, we have the following identity
\be
&\frac{1}{(z+m)^{\Delta_\phi}(\bar z+m)^{\Delta_\phi}} + \frac{1}{(z-m)^{\Delta_\phi}(\bar z-m)^{\Delta_\phi}} \nn \\
&\qquad={1 \over \pi} \int_{0}^{{\sqrt{z \bar z} \over m} } {d w \over w} \left( {1 \over 1 -  w^2 {\bar z \over z}} + {1 \over 1 - w^2 {z \over \bar z}} - 1 \right) {1\over (m - \sqrt{z \bar z} w )^{\Delta_{\phi}} } {2 \sin \pi \Delta_{\phi}  \over ({z \bar z \over w}-m)^{\Delta_{\phi}}} ,
\ee
where we plugged in 
\be
\Disc\left[g\left( {\sqrt{z \bar z} / w} , \sqrt{z \bar{z} }\  w \right) \right]  = \sum_{m=1}^{\infty} {\theta({z \bar z \over w}-m >0) \over (m - \sqrt{z \bar z} w )^{\Delta_{\phi}} } {2 \sin \pi \Delta_{\phi}  \over ({z \bar z \over w}-m)^{\Delta_{\phi}}}
\ee
in the RHS of (\ref{eq:general thermal dispersion relation}). The integral is convergent for $\Delta_{\phi}<1$ and should be defined via a keyhole contour or, equivalently, analytic continuation in $\Delta_{\phi}$ otherwise.

Let us emphasize the following general point neatly illustrated by this example. The Regge limit behavior of $g(z, \bar z)$ and $\Disc [g(z, \bar z)]$ do not have to be the same. Such difference in behaviour is indeed the case here; we have 
\be
g\left( {\sqrt{z \bar z} \over w} , \sqrt{z \bar{z} }\  w \right) \to \frac{1}{z^{\Delta_\phi} \bar z^{\Delta_\phi}}
\ee
in the Regge limit $| w | \to \infty$ and therefore contributes to the first line in (\ref{eq:general thermal dispersion relation}), whereas 
\be
\Disc \left[ g\left( {\sqrt{z \bar z} \over w} , \sqrt{z \bar{z} }\  w \right)\right] \to 0
\ee
which guarantees the convergence of the integral in the second line of (\ref{eq:general thermal dispersion relation}).

\subsection{Connection to the thermal Lorentzian inversion formula}
\label{sec:inversion from dispersion}

Let us discuss the connection of the thermal dispersion relation~(\ref{eq:general thermal dispersion relation}) to the thermal Lorentzian inversion formula~\cite{Iliesiu:2018fao}. The starting point of the thermal Lorentzian inversion formula is to rewrite the OPE in terms of a meromorphic function $a(\De,J)$, with poles
\be
a(\De,J) \sim - \frac{a_\cO}{\De-\De_\cO}\, ,
\ee
by way of the spectral integral
\be
\label{eq:spectral integral}
g(z,\bar z) &= \sum_{J=0}^\infty \oint_{-\e-i\infty}^{-\e+i\infty} \frac{d\De}{2\pi i} a(\De,J) C^{(\nu)}_J\left( \eta \right) r^{\De-2\De_\f} . 
\ee  One also requires that $a(\De,J)$ does not grow exponentially in the $\De$ right-half plane, so that when $r<1$ closing the contour to the right recovers the OPE. The relation~(\ref{eq:spectral integral}) is inverted by way of a Euclidean inversion formula
\be
\label{eq:partialwave}
a(\De,J) = {1 \over \tilde N_J} \int_0^1 {d r \over r} r^{2 \Delta_\phi - \Delta} \int_{-1}^{1} d \eta (1- \eta^2)^{{d-3 \over 2}} C_J^{({d-2 \over 2})}(\eta) g(r,\eta) ,
\ee
where $\tilde N_J$ is the normalization constant
\be
\label{eq:Gegenbauer projection}
\int_{-1}^{1} d \eta (1- \eta^2)^{{d-3 \over 2}} C_J^{({d-2 \over 2})}(\eta) C_{J'}^{({d-2 \over 2})}(\eta) = \tilde N_{J} \delta_{J,J'} .
\ee In~\cite{Iliesiu:2018fao}, the thermal Lorentzian inversion formula was written down by analytically continuing to Lorentzian signature, thereby deforming the $\eta$ contour in~(\ref{eq:partialwave}). 

One can derive the Lorentzian inversion formula by inserting the dispersion relation~(\ref{eq:general thermal dispersion relation}) into the Euclidean inversion formula~(\ref{eq:partialwave}). To compute $a(\De,J)$ starting from the dispersion relation we can switch to the $(r, \eta)$ variables in (\ref{eq:general thermal dispersion relation}) and project onto a given spin by using~(\ref{eq:Gegenbauer projection}). The only nontrivial dependence on $\eta$ in the RHS of (\ref{eq:general thermal dispersion relation}) comes from the factor inside the brackets. Integrating this factor against the Gegenbauer polynomial we get
\be
\label{eq:identityinv}
\int_{-1}^{1} d \eta (1- \eta^2)^{{d-3 \over 2}} C_J^{({d-2 \over 2})}(\eta) {1 - w^4 \over (1+w^2)^2 - 4 w^2 \eta^2} =(1 + (-1)^J ) 2 \pi  \tilde N_J K_J \left(w^{-1} - w\right)^{d-2} F_J(w) 
\ee
whenever $|w|<1$, where
\be
F_J(w) = w^{J+d-2} \ _2 F_1 (J+d-2, {d \over 2} -1 , J+{d \over 2}, w^2), ~~~K_J={\Gamma(J+1) \Gamma({d-2 \over 2}) \over 4 \pi \Gamma(J+{d-2 \over 2})} \, .
\ee

Plugging the dispersion representation of the correlator (\ref{eq:general thermal dispersion relation}) into (\ref{eq:partialwave}), and using (\ref{eq:identityinv}) we get that
\be
\label{eq:TIFL}
a(\De,J) &= (1 + (-1)^J ) \int_0^1 {d r \over r}  \  r^{2 \Delta_\phi - \Delta} {\rm Res}_{w = 0} {2 \pi K_J \over  w}  ({1 \over w} - w)^{d-2} F_J(w)  g \left( {r \over w} , r \  w \right) \nn \\
&\quad + (1 + (-1)^J ) 2 K_J \int_0^1 {d r \over r}  \  r^{2 \Delta_\phi - \Delta} \int_{0}^{r} {d w \over w} \left( {1 \over w} - w \right)^{d-2} F_J(w) \Disc[g\left( {r \over w} , r \  w \right) ] \, .
\ee
This formula precisely agrees with the one in \cite{Iliesiu:2018fao} where the first line corresponds to the arc contribution in the language of  \cite{Iliesiu:2018fao}. 

As mentioned above this formula can be applied as long as for $J>J_0$, where $J_0$ controls the Regge limit of $\Disc[g\left( {r / w} , r w \right)] \sim w^{J_0}$. Note that in situations where $\Disc[g  \left( {r / w} , r  w \right)]$ decays faster in the Regge limit than $g \left( {r / w} , r  w \right)$ the formula above can receive a nontrivial contribution from the first line, see section \ref{subsec:disprelMFT} for a simple example of this type.

\subsection{Subtractions}
\label{subsec{subtractions}}

Above, we considered dispersion relations and their relation to the thermal Lorentzian inversion formula. Let us now combine the two to write down thermal dispersion relations with subtractions, which are applicable to correlators that grow no faster than $|w|^{J_0}$ --- with arbitrary $J_0>0$ --- in the Regge limit. To derive such dispersion relations, we proceed by writing
\be
g(\eta, r) &=\sum_{\Delta} \sum_{\ell=0}^{J_0} a_{\Delta,\ell} C_{\ell}^{({d-2 \over 2})}(\eta) r^{\Delta - 2 \Delta_{\phi}} + \sum_{J=J_0+1}^\infty \oint_{-\e-i\infty}^{-\e+i\infty} \frac{d\De}{2\pi i} a(\De,J) C^{({d-2 \over 2})}_J\left( \eta \right) r^{\De-2\De_\f} ,
\ee
where we explicitly separated the contribution of operators with low spin. We now plug (\ref{eq:TIFL}) for $a(\De,J)$,\footnote{We label by $(\eta', r')$ variables that enter into the integral (\ref{eq:TIFL}) in this section. We hope that it will not create any confusion.} and use the identity (\ref{eq:identityinv}) to write an integral representation for $({1 \over w} - w)^{d-2} F_J(w)$. After this it is convenient to perform the sum using the completeness relation  for Gegenbauer polynomials
\be
\label{eq:identitygeg}
\sum_{J=J_0+1}^\infty {1 \over \tilde N_{J}} (1- \eta'^2)^{{d-3 \over 2}} C_J^{({d-2 \over 2})}(\eta') C_{J}^{({d-2 \over 2})}(\eta)  = \delta(\eta - \eta') - \sum_{J=0}^{J_0} {1 \over \tilde N_{J}} (1- \eta'^2)^{{d-3 \over 2}} C_J^{({d-2 \over 2})}(\eta') C_{J}^{({d-2 \over 2})}(\eta) .
\ee
To accommodate for the fact that only even spin operators contribute in our case, notice that inserting $(-1)^J$ in the formula above is equivalent to the substitution $\eta \to - \eta$.
The $\Delta$ integral is trivially performed using the following identity
\be
\oint_{-\e-i\infty}^{-\e+i\infty} \frac{d\De}{2\pi i} \left({r \over r'} \right)^{\Delta - 2 \Delta_{\phi}} =r' \delta(r-r') ,
\ee
which allows us to easily perform the $r'$ integral in the RHS of (\ref{eq:TIFL}).

In this way we arrive at the following representation for the correlator
\be
\label{eq:subtracteddisp}
g(\eta, r) &=\sum_{\Delta} \sum_{\ell=0}^{J_0} a_{\Delta,\ell} C_{\ell}^{({d-2 \over 2})}(\eta) r^{\Delta - 2 \Delta_{\phi}} + {1 \over \pi} \int_0^{r} {d w \over w} K_{J_0}(w, \eta)   \Disc[g\left( {r \over w} , r \  w \right) ] ,
\ee
where the subtracted kernel $K_{J_0}(w, \eta)$ takes the form 
\be
\label{eq:kernelsub}
K_{J_0}(w, \eta) =  {1 - w^4 \over (1+w^2)^2 - 4 w^2 \eta^2} - 2 \pi \left( {1 \over w} - w \right)^{d-2} \sum_{J=0}^{J_0} (1 + (-1)^J ) K_J C_{J}^{({d-2 \over 2})}(\eta) F_J(w) .
\ee
To derive (\ref{eq:kernelsub}) we used both (\ref{eq:identitygeg}) and then (\ref{eq:identityinv}) once again.

One can check that the Regge behavior of the kernel is improved, as expected. Concretely, we have
\be
\lim_{w \to 0} K_{J_0}(w, \eta) \sim w^{J_0+2} .
\ee

It would be interesting to study dispersive functionals that one gets by imposing the crossing equation, $g(z, \bar z) = g(1-z, 1-\bar z)$, where for $g(z, \bar z)$ we use the formula (\ref{eq:subtracteddisp}). Recently, such dispersive functionals were studied in the context of the vacuum four-point function \cite{Mazac:2019shk,Carmi:2019cub,Penedones:2019tng,Caron-Huot:2020adz}.

\section{On the uniqueness of the solutions}
\label{sec:uniqueness}

One can ask to what extent the solutions to the thermal bootstrap (\ref{Lprop}) are unique given our assumptions. The relevant question
can be restated as follows.  Let us imagine that there exist two different solutions, $g_1(z, \bar z)$ and $g_2(z, \bar z)$, to thermal bootstrap that satisfy our assumptions; in particular, they have only double-trace operators in the OPE and they correctly reproduce the same set of $T=0$ anomalous dimensions $\gamma_{ \ell \leq L}(n)$. However, the two solutions are allowed to have different corrections to the thermal one-point functions $a_{n,\ell}^{(1)}$. Is this possible?

Let us now consider the difference between the two solutions, $\delta g(z, \bar z) \equiv g_1(z, \bar z) - g_2(z, \bar z)$. From our assumptions it follows that it admits the following OPE:
\be
\label{eq:MFT OPE diff}
\delta g(z,\bar z) = \sum_{n=0}^\infty \sum_{\ell=0,2,\dots}^\infty \delta a^{(1)}_{[\f\f]_{n,\ell}} C_{\ell}^{(\nu)}(\eta) (z \bar z)^{n+\ell/2} \, ,
\ee
and similarly in every other OPE channel that can be obtained from (\ref{eq:MFT OPE diff}) by KMS translations.  It also satisfies all the requirements listed in section~\ref{sec:consistency conditions}.

\subsection{Finite number of spins}

Let us first consider a simplified version of this problem where the sum over spins in (\ref{eq:MFT OPE diff}) is bounded by some maximal spin $J_0$ which then automatically coincides with the Regge growth of the correlator. In other words, we would like to exclude the following possibility:
\be
\label{eq:MFT OPE diff J}
\delta g(z,\bar z) &= \sum_{n=0}^\infty \sum_{\ell=0,2,\dots}^{J_0} \delta a^{(1)}_{[\f\f]_{n,\ell}} C_{\ell}^{(\nu)}(\eta) (z \bar z)^{n+\ell/2} \, \nn \\
&= \sum_{\ell=0,2,\dots}^{J_0} C_{\ell}^{(\nu)}(\eta) \delta f_{\ell}(z \bar z) ,
\ee 
where in the second line we introduced unknown functions $\delta f_{\ell}(z \bar z)$. When $| z \bar z | < 1$, each $f_{\ell}(z \bar z)$ is given by the convergent OPE in the first line of (\ref{eq:MFT OPE diff J}), and they are some unknown functions otherwise. 

First, given (\ref{eq:MFT OPE diff J}) it is easy to see that clustering at large spatial separations implies that
\be
\label{eq:clustering}
\lim_{x \to \infty} f_{\ell}(x) = 0 .
\ee
To see this, first notice that large spatial separations correspond to $z \to i \infty$, $\bar z \to - i \infty$, which is the same as $\eta \to 0$ in (\ref{eq:MFT OPE diff J}). To argue for (\ref{eq:clustering}), we need a more general limit where $z = r e^{i \theta}$, $\bar z = r e^{- i \theta}$, $\eta = \cos \theta$ and $r \to \infty$ with fixed $\theta$. To relate it to the spatial clustering we can use KMS invariance to write
\be
\label{eq:clusteringext}
\lim_{r \to \infty} g(r e^{i \theta}, r e^{- i \theta}) = g(r e^{i \theta} - [r \cos \theta], r e^{- i \theta} - [r \cos \theta]) \to 0 ,
\ee
where $[x]$ stands for the integer part of $x$. Note that this argument requires that $\theta$ is finite and $\theta \neq 0, \pi$ (which is necessary for ${\rm Im}[z] = - {\rm Im}[ \bar z] = r \sin \theta \to \infty$ as $r \to \infty$). Applying (\ref{eq:clusteringext}) to $\delta g(z, \bar z)$ and using the fact that the sum over spins in (\ref{eq:MFT OPE diff J}) is finite we get (\ref{eq:clustering}).

To finish the argument, we note that (\ref{eq:clustering}) is inconsistent with KMS translation invariance
\be
\label{eq:KMSdeltag}
\delta g(z,\bar z) &= \delta g(z + m,\bar z + m) \nn \\
&=  \sum_{\ell=0,2,\dots}^{J_0} C_{\ell}^{(\nu)} \left({1 \over 2} \sqrt{z+ m \over \bar z + m} + {1 \over 2} \sqrt{\bar z+ m \over z + m} \right) \delta f_{\ell}((z+m)(\bar z+m)) .
\ee
Indeed, by taking the large $m \to \infty$ limit in the second line of (\ref{eq:KMSdeltag}) and using (\ref{eq:clustering}) we conclude that
\be
\delta g(z, \bar z) = 0 .
\ee
Note that by relaxing clustering, e.g. allowing for non-trivial $\< \phi \>$ so that $f_{\ell=0}(\infty) = \< \phi \>^2$, and applying the same argument at the level of the KMS condition leads to
\be
\delta g(z, \bar z) =  \< \phi \>^2 .
\ee
Therefore, as discussed before, this is a genuine ambiguity of our bootstrap procedure in the absence of $\mathbb{Z}_2$ symmetry.

\subsection{Infinite number of spins}

Let us now say a few words about the case of unbounded spin (\ref{eq:MFT OPE diff}). In this case the argument above fails. While (\ref{eq:clusteringext}) still holds, we cannot derive (\ref{eq:clustering}) from it. To reduce this problem to the previous case of (\ref{eq:MFT OPE diff J})  we need to argue that 
\be
\label{eq:disczerocondition}
{\rm Disc} [\delta g(z,\bar z)] = 0, ~~~~ 0 \le \bar z \le 1 , ~~ 1 \le z . 
\ee
Indeed, given (\ref{eq:disczerocondition}) and Regge boundedness of the thermal correlator we can use the subtracted thermal dispersion relations  (\ref{eq:subtracteddisp}) derived above to conclude that the difference between the two functions is of the type (\ref{eq:MFT OPE diff J}) and apply the previous argument. 

Unfortunately, (\ref{eq:disczerocondition}) does not immediately follow from (\ref{eq:MFT OPE diff}) or its KMS images. We can use the OPE in the $|1-z| , | 1- \bar z| < 1$ channel and the fact that (\ref{eq:MFT OPE diff J}) is analytic in $z$ in that region to establish that
\be
\label{eq:disczeroconditionweak}
{\rm Disc} [\delta g(z,\bar z)] = 0, ~~~~ 0 < \bar z \le 1 , ~~ 1 \le z < 2 ,
\ee
which does not cover the remaining region $z \geq 2$ needed to establish (\ref{eq:disczerocondition}). We do not know how to argue that (\ref{eq:disczeroconditionweak}) continues to hold for $z \geq 2$ and therefore we cannot argue that the solution is unique along these lines. 

The function $\delta g(z, \bar z)$ satisfies all the properties listed in section~\ref{sec:consistency conditions}. Moreover it has a region of zero discontinuity (\ref{eq:disczeroconditionweak}). It is very easy to write functions of this type based on the discussion in the present paper. The simplest example to consider is
\be
\label{eq:counterex}
\delta g(z,\bar z) = \sum_{m = - \infty}^{\infty} {1 \over ( c_0 +(z-m)(\bar z - m))^{\Delta}} ,
\ee
where $c_0 \geq 1$. While this function has all the properties listed in section \ref{sec:consistency conditions} we believe it is still not a viable candidate for the correlator at hand because of its analytic properties in the $(z, \bar z)$-plane. In particular, note that given a fixed $\bar z$, the solutions (\ref{Lprop}) have only singularities at $z = m$, whereas the correlator (\ref{eq:counterex}) has singularities in $z$ that depend on $\bar z$. The singularities at $z=m$ have a natural interpretation as the winding lightcone cuts in the Lorentzian Kaluza-Klein geometry $S^1_\b\times \R^{1,d-2}$ discussed in~\cite{Iliesiu:2018fao}; from this picture, it is unnatural to expect singularities appearing at other locations. It is tempting to conjecture that thermal correlators satisfy this ``extended analyticity'' condition, stating that the only discontinuities are encountered at $z=m$.

It is, however, still clear that assuming this extended notion of analyticity is not strong enough to exclude possible solutions of the type (\ref{eq:MFT OPE diff}). Consider for example the following ansatz:
\be
\label{eq:counterexB}
\delta g(z,\bar z) = \oint_{\e-i\infty}^{\e+i\infty} \frac{ds}{2\pi i} \Gamma(s) \Gamma(\De-s)^2 \delta M_{\beta}(s) G_s(x) ,
\ee
where note that compared to (\ref{Lprop}) we have only a single power of $\Gamma(s)$. As before we assume that $\delta M_{\beta}(s)$ is analytic for ${\rm Re}\ s \geq 0$ and polynomially bounded for large ${\rm Im} \ s$. The fact that there is a single $\Gamma(s)$ in (\ref{eq:counterexB})  guarantees that there are no anomalous dimensions in its OPE. We do not know how to exclude corrections of the type (\ref{eq:counterexB}) based on general principles, including extended analyticity, and we leave a better understanding of this issue to future study.\footnote{Note that, as explained at $T=0$ in~\cite{Penedones:2019tng}, different choices of $\Gamma$-function prefactors in (\ref{eq:counterexB}) lead to different analytic properties of the correlation function (assuming $ \delta M_{\beta}(s)$ is polynomially bounded). It is therefore tempting to think that the physical solutions (\ref{Lprop}) could be selected over (\ref{eq:counterexB}) based on a better understanding of the analytic properties of the physical thermal correlator.}

\section{Discussion and conclusions}
\label{sec:conclusion}

In this paper we have presented an infinite number of corrections to the MFT thermal two-point function, corresponding to quartic vertices in the bulk with an arbitrary number of derivatives. The input for our proposal are the anomalous dimensions for the intermediate double-trace operators, which can be obtained from a crossing problem at zero temperature. We have shown that our proposal satisfies all consistency conditions. We have also derived a dispersion relation for thermal two-point functions. There are several directions that would be interesting to explore. 

\begin{itemize}
\item It would be interesting to consider solutions corresponding to exchange diagrams.\footnote{At zero temperature, the analogous question was analyzed in \cite{Alday:2017gde}.} In this case we will have a new operator in the OPE, and operators of arbitrarily high spin acquire anomalous dimensions, which are again fixed by a crossing problem at zero temperature. The corrections can be determined either from consistency conditions, or via thermal Witten diagrams. 
\item It would be interesting to prove uniqueness, or otherwise systematically understand what are the possible ambiguities in the solutions to the thermal bootstrap.  We believe that a better understanding of the analytic properties of the correlator will be crucial to achieve this goal.
\item Our answer is suggestive of a formulation of thermal Mellin amplitudes. It would be interesting to make this more precise. 
\item It would be interesting to study the presence and properties of Landau equations/poles, and the bulk point singularity in the thermal setting \cite{Gary:2009ae,Maldacena:2015iua,Fitzpatrick:2016ive,DodelsonOoguri}. The tower of solutions we have found are the arena to start this exploration. 
\item The solutions that we found in this paper are given in terms of meromorphic functions $M_{\beta,\ell}(s)$, which have many interesting properties that we have not fully explored. They are curiously related to analytic continuation of the underlying OPE data of double trace operators $[\f\f]_{n,\ell}$ in $n$. We also have not explored the origin of the singularities of $M_{\beta,\ell}(s)$ that control the large spatial separation expansion of the correlator. Additionally, we note that the $M_{\beta,\ell}(s)$ that we found are polynomially bounded after dividing by a $\zeta$-function of an appropriate argument; therefore, they admit dispersion relations in $s$ which are separate from the dispersion relations studied in section \ref{sec:thermal dispersion relation}. 
\item In this paper we have considered conformal correlators dual to dynamics in thermal AdS. In an honest CFT, the presence of the stress tensor modifies the bulk phase to AdS black hole or black brane spacetimes. It would be very interesting to consider CFTs dual to AdS black holes. In future work, we hope to apply the approach we developed to study thermal correlators in black hole backgrounds.
\end{itemize}

Let us make a small comment about the black hole case, leaving a more detailed discussion for future work. Considering quartic corrections to the potential (\ref{eq:perturbative QFT in AdS})  leads to two different effects in this case. One is the correction of the double-trace OPE and thermal one-point function data. The second is the correction to the OPE data of the multi-trace stress tensor operators which are also present in the OPE in this case. We do not have good control over the latter, see e.g. \cite{Kulaxizi:2018dxo,Fitzpatrick:2019zqz,Li:2019zba} for some recent progress in this direction. However, nothing stops us from writing the following formula that accommodates for the anomalous dimensions of the double-trace operators
\be
\label{eq:ansatzBH}
\< \phi(z,\bar z)\phi(0)\>^{\text{Double-trace}}_{\text{BH}} |_{\lambda_i^{1}} \overset{?}{=} \oint_{\e-i\infty}^{\e+i\infty} \frac{ds}{2\pi i} \Gamma(s)^2 \Gamma(\De-s)^2 M_{\text{BH}}^{\text{DT}}(s) \< \phi(z,\bar z)\phi(0)\>_\b^{\text{MFT,} \De_\f=s} ,
\ee
where the thermal Mellin amplitude $M^{\text{DT}}_{\text{BH}}(s)$ is fixed by consistency with the OPE as before. The crucial difference is that this time the relevant zeroth-order thermal coefficients $a^{\text{BH}}_{[\f\f]_{n,\ell}}$ are not known explicitly. Assuming that these can be computed and analytically continued in $n$ such that the resulting function $M^{\text{DT}}_{\text{BH}}(s)$ and the Mellin integral above are properly convergent, we get that the formula above satisfies all the expected properties. This is yet another illustration of something that our analysis hopefully made clear: requiring consistency of the thermal two-point function is not enough to fix it, and the input of the vacuum OPE data (which is fixed by solving crossing equations in vacuum) is crucial to find the finite temperature answer. 

\section*{Acknowledgements} 
We would like to thank A. Bissi, Y. Jiang, B. Mukhametzhanov, and E. Perlmutter for discussions on related issues. This project has received funding from the European Research Council (ERC) under the European Union's Horizon 2020 research and innovation programme (grant agreement No 787185). 

\appendix

\section{Anomalous dimensions of double-trace operators}
\label{app:anomalous dimensions}
In this appendix we collect anomalous dimensions for double-trace operators for various vertices. This serves as an input to construct the thermal correlators considered in this paper.  Let us start with the quartic vertex with no derivatives $\Phi^4$. Only double-trace operators with spin zero acquire an anomalous dimension, which we will denote by $\gamma^{(0)}_{n,0}$. For $d=2,4$ this was computed in~\cite{Heemskerk:2009pn}, while for general $d$ it was computed in~\cite{Fitzpatrick:2010zm}. It is given by 
\begin{equation}
\label{eq:spin 0 anomalous dimensions}
\gamma^{(0)}_{n,0}= \frac{2^{-d-2} \pi ^{-\frac{d}{2}} \Gamma \left(\frac{d}{2}+n\right) \Gamma (n+\Delta ) \Gamma \left(-\frac{d}{2}+n+\Delta +\frac{1}{2}\right) \Gamma \left(-\frac{d}{2}+n+2 \Delta \right)}{\Gamma \left(\frac{d}{2}\right) \Gamma (n+1) \Gamma \left(n+\Delta +\frac{1}{2}\right) \Gamma \left(-\frac{d}{2}+n+\Delta +1\right) \Gamma (-d+n+2 \Delta +1)} \, .
\end{equation}
Next, we focus in interactions with four derivatives, schematically of the form $\left( \nabla \Phi \cdot \nabla \Phi  \right)^2$. In this case, double-trace operators of spin $\ell=0,2$ acquire anomalous dimension, which we will denote by $\gamma^{(2)}_{n,\ell}$. The simplest way to compute them is to focus in the corresponding crossing symmetric four-point correlator of external operators of dimension $\Delta$ and perform the conformal block expansion. As shown in \cite{Heemskerk:2009pn}, this is given by
\begin{equation}
{\cal G}_{\Delta}(u,v) = u^{\Delta} (1+u +v) {\bar D}_{\Delta+1,\Delta+1,\Delta+1,\Delta+1}(u,v)
\end{equation}
where ${\bar D}-$functions have been defined, for instance, in~\cite{Dolan:2000ut}. From their results we can compute the piece proportional to $\log u$ for the corresponding ${\bar D}-$function, and find
\begin{equation}
\left. {\bar D}_{\Delta+1,\Delta+1,\Delta+1,\Delta+1}(u,v) \right|_{\log u} = - \sum_{m,n=0} \frac{\Gamma (m+\Delta+1)^2 \Gamma (m+n+\Delta+1)^2}{\Gamma (m+1)^2 \Gamma (n+1) \Gamma (2 m+n+2 \Delta+2)} u^m (1-v)^n 
\end{equation}
We can then perform the conformal block decomposition of the corresponding contribution in ${\cal G}_{\Delta}(u,v)$. For $d=2,4$ the MFT OPE coefficients, as well as the expression for the conformal blocks can be found in \cite{Heemskerk:2009pn}. In $d=2$ we find
\begin{eqnarray}
\gamma^{(2)}_{n,0} &=& \frac{P^{2d}_6(n)}{(2 \Delta +2 n-3) (2 \Delta +2 n-1) (2 \Delta +2 n+1)},\\
\gamma^{(2)}_{n,2} &=& \frac{(n+1)_2 (\Delta +n)_2 (2 \Delta +n-1)_2}{(2 \Delta +2 n-1) (2 \Delta +2 n+1) (2 \Delta +2 n+3)},
\end{eqnarray}
where $P^{4d}_6(n)$ is a degree six polynomial given by
\begin{eqnarray*}
P^{2d}_6(n)&=&4 \Delta ^4 \left(4 \Delta ^2-8 \Delta +3\right)+4 \Delta ^3 \left(16 \Delta ^2-26 \Delta +9\right) n +\\
& & +2 \left(76 \Delta ^4-114 \Delta ^3+31 \Delta ^2+11 \Delta -5\right) n^2 + \left(232 \Delta ^3-300 \Delta ^2+80 \Delta +6\right) n^3 + \\
& & +2 \left(99 \Delta ^2-93 \Delta +16\right) n^4 +(84 \Delta -42) n^5+14 n^6.
\end{eqnarray*}
In $d=4$ we find
\begin{eqnarray}
\gamma^{(2)}_{n,0} &=& \frac{(n+1) (\Delta +n-1) (2 \Delta +n-3) P^{4d}_6(n)}{4 (2 \Delta +2 n-5) (2 \Delta +2 n-3) (2 \Delta +2 n-1) (2 \Delta +2 n+1)}, \\
\gamma^{(2)}_{n,2} &=& \frac{(n+1)_3 (n+2 \Delta -3)_3 (n+\Delta -1)_3}{12 (2 \Delta +2 n-3) (2 \Delta +2 n-1) (2 \Delta +2 n+1) (2 \Delta +2 n+3)},
\end{eqnarray}
where $P^{4d}_6(n)$ is a degree six polynomial given by
\begin{eqnarray*}
P^{4d}_6(n)&=&16 \Delta ^6-64 \Delta ^5+68 \Delta ^4-20 \Delta ^3+ \left(64 \Delta ^5-216 \Delta ^4+212 \Delta ^3-60 \Delta ^2\right) n \\
&&+ \left(148 \Delta ^4-460 \Delta ^3+394 \Delta ^2-40 \Delta -56\right) n^2 + \left(220 \Delta ^3-580 \Delta ^2+408 \Delta -48\right) n^3 \\
& &+ \left(185 \Delta ^2-350 \Delta +142\right) n^4+(78 \Delta -78) n^5+ 13 n^6.
\end{eqnarray*}
Finally, we focus in the interaction with six derivatives that truncates at spin two, schematically of the form $(\nabla \Phi)^2 \Phi (\nabla_\mu \nabla_\nu \Phi)^2$. Again, the four point function is given in HPPS in terms of ${\bar D}-$functions. We can perform the conformal block decomposition and read off the anomalous dimension. We will only be interested in the anomalous dimension for spin two operators. For $d=2$ we obtain
\begin{eqnarray}
\gamma^{(2')}_{n,2} &=&\frac{(n+1)_2 (\Delta +n)_2 (2 \Delta +n-1)_2 \left(6 \Delta ^2+7 \Delta +3 n^2+(6 \Delta +3) n+2\right)}{(2 \Delta +2 n-1) (2 \Delta +2 n+1) (2 \Delta +2 n+3)},~~~\text{for $d=2$}, \nonumber\\
\gamma^{(2')}_{n,2} &=&\frac{(n+1)_3 (\Delta +n-1)_3 (2 \Delta +n-3)_3 \left(2 \Delta  (3 \Delta +2)+3 n^2+6 \Delta  n\right)}{(2 \Delta +2 n-3) (2 \Delta +2 n-1) (2 \Delta +2 n+1) (2 \Delta +2 n+3)},~~~\text{for $d=4$},
\end{eqnarray}
where some convenient normalisation has been chosen.

\bibliographystyle{JHEP}
\bibliography{refs}

\end{document}